\documentclass[12pt]{article}
\usepackage{ytableau}
\usepackage{tikz, tikz-cd}
\usepackage{savesym}
\usepackage{cite}
\usepackage{wasysym}
\usepackage{amscd}
\usepackage{graphicx}
\usepackage{pifont}
\usepackage{float}
\usepackage{subfig}
\usepackage[all]{xy}
\usepackage{multicol}
\usepackage{amsfonts}
\usepackage{picture}
\usepackage{amssymb}
\savesymbol{iint}
\savesymbol{iiint}
\usepackage{amsmath}
\restoresymbol{TXF}{iint}
\restoresymbol{TXF}{iiint}
\usepackage{color}
\usepackage{clay}
\usepackage{hyperref}
\usepackage{slashed}
\hypersetup{colorlinks=true}
\hypersetup{linkcolor=black}
\hypersetup{citecolor=black}
\hypersetup{urlcolor=black}
\usepackage{setspace}
\usepackage{multirow}
\usepackage{wrapfig}
\usepackage{verbatim}
\numberwithin{equation}{section}

\begin{document}

\date{June, 2014}

\institution{Fellows}{\centerline{${}^{1}$Society of Fellows, Harvard University, Cambridge, MA, USA}}
\institution{HarvardU}{\centerline{${}^{2}$Jefferson Physical Laboratory, Harvard University, Cambridge, MA, USA}}

\title{An Index Formula for Supersymmetric Quantum Mechanics}

\authors{Clay C\'{o}rdova\worksat{\Fellows}\footnote{e-mail: {\tt cordova@physics.harvard.edu}}  and Shu-Heng Shao\worksat{\HarvardU}\footnote{e-mail: {\tt shshao@physics.harvard.edu}} }

\abstract{We derive a localization formula for the refined index of gauged quantum mechanics with four supercharges.  Our answer takes the form of a residue integral on the complexified Cartan subalgebra of the gauge group.  The formula captures the dependence of the index on Fayet-Iliopoulos parameters and the presence of a generic superpotential.  The residue formula provides an efficient method for computing cohomology of quiver moduli spaces.  Our result has broad applications to the counting of BPS states in four-dimensional $\mathcal{N}=2$ systems.  In that context, the wall-crossing phenomenon appears as discontinuities in the value of the residue integral as the integration contour is varied.  We present several examples illustrating the various aspects of the index formula.}

\maketitle
\tableofcontents
\section{Introduction}

Supersymmetric quantum mechanics has a wide variety of applications in mathematical physics.  It arises universally as the zero momentum sector of supersymmetric field theories and governs the worldline dynamics of supersymmetric particles.  A basic feature of any such system is its set of supersymmetric ground states.  When these states are counted with signs according to their fermion number they form the Witten index \cite{Witten:1982df}, perhaps the most primitive example of a quantity protected by supersymmetry.

Motivated by these general considerations, in this work we determine a general formula for the index of $\mathcal{N}=4$ quantum mechanics.    We focus on the class of quantum mechanics models that have Lagrangians which arise from the dimensional reduction of four-dimensional supersymmetric gauge theories.  In this context the counting of vacua may be further sharpened using $R$-charges.  The result is a refined index
\begin{equation}
\Omega \equiv \mathrm{Tr}_{\mathcal{H}}\left(\phantom{\int}\hspace{-.16in}(-1)^{F}\exp(-\beta H)y^{R+2J_{3}}\right).
\end{equation}
Our main result is an integral expression for $\Omega$ derived by supersymmetric localization \cite{Witten:1988ze, Pestun:2007rz}.

Pragmatically speaking, our derivation of the index formula in \S \ref{index} follows closely a similar calculation for the elliptic genus of two-dimensional systems with $\mathcal{N}=(2,2)$ supersymmetry.  Consequently, our final answer for the index $\Omega$ takes a similar form to that uncovered in \cite{Gadde:2013dda, Benini:2013nda, Benini:2013xpa}:  the index $\Omega$ can be expressed as a residue integral of a meromorphic form on a product of complex annuli $(\mathbb{C}^{*})^{r}.$

The index $\Omega$ depends in a subtle way on two pieces of data entering the quantum-mechanical model.
\begin{itemize}
\item In gauge theories with abelian factors, the Lagrangian may contain Fayet-Iliopoulos parameters $\zeta.$  The index $\Omega$ depends in a piecewise constant fashion on such FI parameters.  Across codimension one walls in $\zeta$-space, supersymmetric vacua may be created or destroyed and the index $\Omega$ jumps.  In our context, the FI parameters enter the index through a specification of integration contour.  The jumping of the index is mapped to the change of a residue integral under large variations in the contour. 

\item In theories which admit non-trivial superpotentials, the refined index $\Omega$ depends on the superpotential through the $R$-charge assignments that the latter implies for chiral fields.  We find that the residue formula accurately encodes this dependence for the case of generic superpotential.
\end{itemize}
We highlight these key features of the index in our study of examples in \S \ref{examples}.

In \S\ref{cohomology} we compare the residue formula to alternative computational approaches to the index.  The most straightforward technique involves two steps. First, one calculates the classical moduli space of the supersymmetric quantum mechanics.  Then, one finds the desired ground state wavefunctions by quantizing the moduli space, i.e. computing its cohomology.  Our residue formula bypasses the intermediate step of the classical moduli space and computes directly the refined index $\Omega$ which may be interpreted as a generating functional of the cohomology.  In this way our index formula is similar in spirit to the Reineke formula \cite{2003InMat.152..349R} for the cohomology of moduli spaces of quiver representations, and to its cousin the MPS formula \cite{Manschot:2011xc} obtained by geometric quantization of the Coulomb branch.

One of the key physical applications of the index formula occurs in the study of BPS states in four-dimensional systems with $\mathcal{N}=2$ supersymmetry.  Often, the BPS spectrum may be described via the ground states of quiver quantum mechanics.  We briefly review this connection in \S \ref{BPSStates}. The class of physical systems to which this paradigm applies is broad and includes black holes in supergravity \cite{Douglas:2000ah, Douglas:2000qw, Denef:2002ru, Manschot:2012rx}, dyons in four-dimensional gauge theories \cite{Fiol:2000pd, Alim:2011kw, Cecotti:2012gh}, and even more exotic systems decorated by external defects \cite{Chuang:2013wt, Cordova:2013bza}.

In the context of BPS states, our result for the quantum mechanical index $\Omega$ can be interpreted as an explicit formula for the protected spin character of BPS states with an electromagnetic charge determined by the ranks of the quiver gauge groups.  The jumps in $\Omega$ as the FI parameters are varied are then mapped to the ubiquitous wall-crossing phenomenon first uncovered in \cite{Cecotti:1992rm, Seiberg:1994rs, Seiberg:1994aj}.  The fact that wall-crossing may be encoded by contour deformation of a residue integral is a generalization of similar ideas in systems with $\mathcal{N}=4$ supersymmetry \cite{Cheng:2007ch}.  

Wall-crossing has recently been extensively studied \cite{Gaiotto:2008cd,  Gaiotto:2009hg, Cecotti:2009uf, Dimofte:2009tm,  Manschot:2010qz, Kim:2011sc} due to the existence of universal formulas \cite{Denef:2007vg, Kontsevich:2008fj, Joyce:2008pc} encoding the discontinuities in the BPS spectrum.  In the simple examples that we have investigated, the discontinuities in the residue formula for $\Omega$ agree with these universal formulas.  It would be interesting to understand the relation more concretely and explain why our residue prescriptions obey wall-crossing formulas.  We leave this, as well as applications of the index formula to interesting four-dimensional $\mathcal{N}=2$ systems, as open problems for future work.

\textit{Note added}: While this work was being completed the preprint \cite{Hwang:2014uwa} appeared which develops the same formula for the refined index in the context of generalized ADHM quantum mechanics.  Localization formulas for the index of supersymmetric quantum mechanics have also been independently obtained in \cite{Lecture1, Hori:2014tda}.  See additionally \cite{ Ohta:2014ria} for related work.

\section{The Index of $\mathcal{N}=4$ Quantum Mechanics}
\label{index}

In this section we present the residue formula for the index of $\mathcal{N}=4$ quantum mechanics.  Our derivation follows straightforwardly from the dimensional reduction of the elliptic genus formulas of \cite{Benini:2013nda, Benini:2013xpa}.  Our discussion is brief and we refer to those works for a more complete treatment.

\subsection{Gauged Quantum Mechanics and the Refined Index}
\label{lagrangians}
The class of models we consider are quantum-mechanical gauge theories with four real supercharges.  We assume throughout that the system is gapped so that there are a finite number of ground states which are separated in energy from the excited states.  Our aim is to count (with appropriate signs), the number of ground states in such a model.

In addition to possible flavor symmetries, the systems in question have $R$-symmetry group $su(2)_{J}\times u(1)_{R}$.  There are two classes of multiplets:
\begin{itemize}
\item Vector multiplets associated to gauge groups.  The bosonic fields consist of a one-component gauge field $A$ and a triplet of adjoint scalars $\vec{X}$.  The gauge field is uncharged under the $R$-symmetry group, while the adjoint scalars transform in a $\mathbf{3}$ of $su(2)_{J}$ and are neutral under $u(1)_{R}$.

\item Chiral multiplets associated to representations of the gauge groups.  The bosonic fields consist of a complex scalar $\Phi$ transforming as a singlet under $su(2)_{J}$ and with $u(1)_{R}$ charge $R_{\Phi}$.
\end{itemize}   

In addition to the spectrum of vector and chiral multiplets the Lagrangian for our quantum mechanics depends on two additional pieces of data.
\begin{itemize}
\item FI parameters.  Let the total gauge symmetry algebra for the quantum mechanics be $\frak{g}$. We decompose $\frak{g} =  \frak{\tilde g}+ \frak{g}_{u(1)},$ where $\frak{\tilde g}$ is semi-simple and $\frak{g}_{u(1)}=\bigoplus_i \frak{u(\text{1})_i} $ is the abelian part of the gauge algebra.  We view the FI parameter $\zeta$ as an element of the dual space $\frak{g}_{u(1)}^{*}$.

\item Superpotentials.  If the model admits holomorphic gauge invariant monomials in the chiral fields then we may activate them in the superpotential $\mathcal{W}.$    Consider a monomial in $\mathcal{W}$ and let $d_{i}$ denote the degree in this monomial of the chiral field $\Phi_{i}$.  The presence of such a term restricts the $R$-charges of the chirals as
\begin{equation}
R(\mathcal{W})=2= \sum_{i}{d}_{i}R_{\Phi_{i}}. \label{degconst}
\end{equation}
The above constraint must be true for each monomial term in the superpotential and restricts $\mathcal{W}$ to be quasi-homogeneous.  

In our analysis, the superpotential will enter only through the above constraints on the $u(1)_{R}$ charges of chiral fields.  Thus our results are restricted to the case of quasi-homogeneous superpotential.  Aside from the constraint \eqref{degconst}, the $u(1)_{R}$ charges of chiral fields may be chosen arbitrarily.  

We make two additional assumptions about $\mathcal{W}$.
\begin{itemize}
\item We assume that all lowest degree terms consistent with quasi-homogeneity are in fact present in $\mathcal{W}$.\footnote{Thus, if a quadratic superpotential is possible we assume that it is present.  If no quadratic superpotential is possible and a cubic potential is possible we assume the later is present.  And so on.}
\item We assume that $\mathcal{W}$ is a generic polynomial of multi-degree consistent with \eqref{degconst} and the previous assumption.
\end{itemize}
As we illustrate in the examples of \S \ref{potentialsec}, both of these assumptions are necessary for the applicability of the residue formula of \S \ref{residuesec}.\footnote{Indeed without these additional assumptions, the spectrum will generally be non-discrete and the index as studied here is incomplete.}
\end{itemize}

Given a fixed gauged quantum mechanics, our object of interest is the refined Witten index defined as
\begin{equation}
 \Omega(y, \zeta) \equiv \mathrm{Tr}_{\mathcal{H}}\left(\phantom{\int}\hspace{-.16in}(-1)^{F}\exp(-\beta H)y^{R+2J_{3}}\right).
\end{equation}
As usual, when the system is gapped the index receives contributions only from ground states and hence is independent of $\beta$. In general, the index depends on both the FI parameter $\zeta$ and the $R$-charges of chiral fields.\footnote{We suppress the dependence on $R$-charges in the notation.}  The charge $R+2J_{3}$ commutes with the supercharge used to form the index and hence we may further grade the ground states to obtain a non-trivial function of $y$.  It is convenient to define $z$ as
\begin{align}
y=e^{i \pi z}.
\end{align}
In the following we use $z$ and $y$ interchangeably.

\subsection{The Residue Formula for the Index}
\label{residuesec}

The refined index $\Omega(y, \zeta)$ can be computed by a path integral on a circle with periodic boundary conditions for fermions, and background $R$-symmetry gauge fields.  A formula $\Omega(y, \zeta)$ can be directly obtained by taking the dimensional reduction limit of the various ingredients of the localized elliptic genus formula of \cite{Benini:2013nda, Benini:2013xpa} (see \S3 of \cite{Benini:2013xpa} for the derivation). Our final answer takes the form of a residue 
\begin{align}
\Omega(y, \zeta) = {1\over |W|} \sum_{u_* \in \frak{M}^*_{\text{sing}}} \underset{u=u_*}{\text{JK-Res}}\left(
\mathbf{Q}(u_*) ,  \zeta  \right)
\, Z_{1-\text{loop}} (z,u), \label{residueform}
\end{align}
where $|W|$ is the order of the Weyl group and $\zeta$ is the FI parameter. 

In this section we explain the elements of this formula.  In the remainder of the paper we discuss its various applications.

\paragraph{Definition of the Space $\frak{M}$}\mbox{}\\

The $u$ variable that appears in \eqref{residueform} is valued in a space $\frak{M}$ of bosonic zero modes of the vector multiplets.  We restrict the gauge field and scalars to be valued in the Cartan subalgebra $\frak{h}$ of the gauge algebra $\frak{g}$. In the triplet of scalars in the vector multiplet, there is one real component which is neutral under the charge $R+2J_{3}$ and we denote this field by $X$.  The field $X$ may have zero modes, while for generic $y$, the remaining members of the triplet do not have zero modes.

The definition of the variable $u$ is then
\begin{align}
u \equiv A^{(0)} - i X^{(0)},
\end{align}
where $A^{(0)}$ and $X^{(0)}$  are the zero modes for the one-dimensional gauge field and the scalar $X$.  Since $A$ is a gauge field, large gauge transformations make the real part of $u$ periodic.  Thus the space $\frak{M}$ of zero modes is a product of annuli
\begin{align}
\frak{M} = \frak{h}_{\mathbb{C}}/Q^\vee\cong(\mathbb{C}^*)^r\end{align}
where $r$ is the total rank of the gauge groups  and $Q^\vee$ is the coroot lattice. 

\paragraph{Definition of the Meromorphic Form $Z_{1-\text{loop}} (z,u)$}\mbox{}\\

The quantity $Z_{1-\text{loop}} (z,u)$ is a meromorphic top form on the space $\frak{M}.$  It is defined by computing the one-loop determinant of the massive modes in the path integral on the circle.  This one-loop determinant receives contributions from the vector multiplets and the chiral multiplets as
\begin{align}
Z_{1-\text{loop}} =  \prod_V Z_{V,G} \prod_\Phi Z_{\Phi,\mathbf{R}} .
\end{align}
The quantities $Z_{V,G}$ and $Z_{\Phi,\mathbf{R}}$ can be obtained from direct dimensional reduction of (2.12) and (2.8) in \cite{Benini:2013nda}, respectively.

The contribution of a vector multiplet $V$ with gauge group $G$ to the one-loop determinant $Z_{1-\text{loop}}$ is
\begin{align}
Z_{V,G}(z,u) = \left[ -{ \pi \over \sin(\pi z) }\right]^{\text{rank} \, G}
\prod_{\alpha \in G } { \sin[\pi \alpha(u)  ] \over \sin[ \pi \alpha(u) - \pi z]}\prod_{a=1}^{\text{rank}\, G} du_a.
\end{align}
where the product of $\alpha$ is over the roots of $G$.

The contribution of a chiral multiplet $\Phi$ in the representation $\mathbf{R}$ with $u(1)_{R}$ charge $R$ is
\begin{align}
Z_{\Phi,\mathbf{R} } (z,u) = \prod_{\rho \in \mathbf{R}} 
{ \sin\left[ \pi \rho(u)+ \pi \left( {R\over 2}-1\right) z\right] \over \sin\left[ \pi\rho(u) +\pi {R\over2} z\right]},
\end{align}
where the product of $\rho$ is over the weights of $\mathbf{R}$.

\newpage
\paragraph{Definition of the Locus $\frak{M}^*_{\text{sing}}$}\mbox{}\\

Next we define the locus  $\frak{M}^*_{\text{sing}}\subset \frak{M}$. The form $Z_{1-\text{loop}}$ has poles along hyperplanes $H_i$ in $\frak{M}$ where modes, which are massive at generic $u,$ become massless.  Specifically these hyperplanes are
\begin{align}
&\text{vector}:~H_i=\Big\{ -z + Q_i(u) =0~~~\text{mod}~\mathbb{Z} \Big\}&Q_i =\alpha,\\
&\text{chiral}:~H_i =\left\{ {R_i\over 2}z +Q_i(u) = 0~~~\text{mod}~\mathbb{Z}\right\}&Q_i=\rho.
\end{align}
And the charge covectors $Q_i \in \frak{h}^*$ can be either the roots $\alpha$ of the gauge algebra or weights $\rho$ of the matter representations. 

We define
\begin{align}
\frak{M}^*_{\text{sing}} = \left\{
u_*\in \frak{M} \,\Big|\, \text{at least $r$ linearly independent $H_i$'s meet at $u_*$}
\right\}.
\end{align}
$\frak{M}_{\text{sing}}^*$ is the collection of points where the residue \eqref{residueform} is evaluated.

\paragraph{Definition of the Residue}\mbox{}\\

The Jeffrey-Kirwan residue operation $\underset{u=u_*}{\text{JK-Res}}\left(
\mathbf{Q}(u_*) ,\eta
\right)$ is defined abstractly in \cite{MR1318878} and studied constructively in \cite{2004InMat.158..453S}.

For notational simplicity, we shift the point where we evaluate the residue to be at $u_*=0$. $\mathbf{Q}(u_*)$ is a collection of charge covectors $Q_i\in \frak{h}^*$ with $i=1,\cdots,n$ for some $n$. The collection $\mathbf{Q}(u_*)$ defines $n$ hyperplanes meeting at $u=u_*$:
\begin{align}
H_i = \left\{ u\in \mathbb{C}^r \,\Big|\, Q_i(u)=0\right\}.
\end{align}
In addition, the Jeffrey-Kirwan residue operation depends on a choice of covector $\eta\in \frak{h}^*$.

If all the charge covectors in $\mathbf{Q}(u_*)$ are contained in a half-space of $\frak{h}^*$, the hyperplane arrangement is said to be projective. For a projective arrangement, the Jeffrey-Kirwan residue is the linear functional defined by the conditions
\begin{align}\label{JK}
\underset{u=u_*}{\text{JK-Res}}\left(
\mathbf{Q}(u_*) ,\eta
\right)
{du_1 \over Q_{j_1}(u)}\wedge\cdots\wedge
{du_r \over Q_{j_r}(u)}
=
\begin{cases}
\det|(Q_{j_1}\cdots Q_{j_r})|^{-1}~~&\text{if} ~\eta\in \text{Cone}(Q_{j_1}\cdots Q_{j_r}),\\
0~~&\text{otherwise},
\end{cases}
\end{align}
where $\text{Cone}(Q_{j_1}\cdots Q_{j_r})$ indicates the positive linear span of the covectors $Q_{j_1},\cdots, Q_{j_r}$. In particular, if $n=r$, the hyperplane arrangement is projective. For simplicity in this paper we study examples with $n=r$.


\paragraph{Definition of the Contour}\mbox{}\\

Finally, we must specify the choice of the covector $\eta\in \frak{h}^*$ in the definition of the Jeffrey-Kirwan residue operation \eqref{JK}. This quantity is fixed by the FI parameter $\zeta$ as
\begin{align}\label{etazeta}
\eta =  \zeta \in \frak{g}_{u(1)}^{*} \subset  \frak{h}^{*}.
\end{align}
This identification may be justified by demanding that the gaussian path integral over the zero mode of the auxiliary $D$ field in the vector multiplet passes through the point of stationary phase fixed by the FI parameter $\zeta$.

Equation \eqref{etazeta} is a key aspect of the residue formula \eqref{residueform}.  Because of the discontinuity in the Jeffrey-Kirwan residue operation \eqref{JK} as $\zeta$ varies, the contour prescription \eqref{etazeta} enables the index $\Omega(y, \zeta)$ to depend in a piecewise constant fashion on the FI parameter.

\subsection{Cohomology of Higgs Branch Moduli Spaces}
\label{cohomology}

It is fruitful to compare the residue formula for the refined index to other methods of calculating the ground states.  

The most direct approach is to calculate the moduli space of classical vacua and then quantize this moduli space to determine wavefunctions.   As usual in supersymmetric gauge theory, the classical moduli space is typically separated into multiple branches: Higgs branches where matter fields are non-vanishing, and Coulomb branches where scalars from the vector multiplets are non-vanishing.  We isolate one of these branches and quantize.  We focus on the Higgs branch as it is typically better behaved.  For Coulomb branch approaches see \cite{Denef:2002ru, Manschot:2011xc, Kim:2011sc}.

The classical Higgs branch $\mathcal{M}$ is simply the set of solutions to the $F$ and $D$ flatness conditions modulo the action of the gauge group. 

Explicitly, let $G$ denote the total gauge group of the gauged quantum mechanics. And let $\Phi_{\nu}$ indicated the chiral fields transforming in representations $\mathbf{R}_{\nu}$ of $G$.  We define a set in the vector space $\oplus_{\nu}\mathbf{R}_{\nu}$ as the set of  $\Phi_{\nu}$ obeying the following equations.
\begin{itemize}
\item For each chiral field $\Phi_{\nu},$ the superpotential is stationary
\begin{equation}
\frac{\partial \mathcal{W}}{\partial\Phi_{\nu}} = 0. \label{Fflat}
\end{equation}
\item The gauge group $G$ has a number of abelian factors, each with an associated FI parameter $\zeta_{i}$. Let $q_{\nu}^{i}$ denote the charge of $\Phi_{\nu}$ under the $i$-th $U(1)$.  Then for each abelian factor we demand
\begin{equation}
\sum_{\nu}q_{\nu}^{i} |\Phi_{\nu}|^{2}= \zeta_{i}. \label{Dflat}
\end{equation}
\end{itemize}
The Higgs branch moduli space $\mathcal{M}$ is the set of solutions to \eqref{Fflat}-\eqref{Dflat} quotiented by the action of the group $G$.

In the most widely studied class of examples, the gauge group $G$ is a product of unitary groups and the representations $\mathbf{R}_{\nu}$ are chosen to be bifundamentals.  In that case $\mathcal{M}$ is the moduli space of stable quiver representations \cite{KING01121994}.  

In favorable circumstances, the moduli space $\mathcal{M}$ is compact and we may now extract the ground state spectrum from its cohomology.  To form the refined index we must then assemble this cohomology into a generating function.  Supersymmetry implies that $\mathcal{M}$ is K\"{a}hler and hence its cohomology may be bigraded into Dolbeault cohomology groups.  We denote by $h^{p,q}\left(\mathcal{M}\right)$ the resulting Hodge numbers, and let $d$ denote the complex dimension of $\mathcal{M}$.  Then the refined index is
\begin{equation}
\Omega(y,\zeta)= \sum_{p,q=0}^{d} h^{p,q}\left(\mathcal{M}\right)(-1)^{p-q}y^{2p-d}. \label{cohomindex}
\end{equation}

Agreement between \eqref{cohomindex} and the residue formula \eqref{residueform} yields a direct way of extracting information about the cohomology of the moduli space $\mathcal{M}$ which is similar in spirit to \cite{2003InMat.152..349R}.  Note however that the residue formula  \eqref{residueform} is applicable only in the case of discrete spectrum which in the context of quiver representations implies that ranks of the gauge groups must be coprime.

\section{Relation to BPS Particles of $4d$ $\mathcal{N}=2$ Systems}
\label{BPSStates}

In this section, we briefly review the connection between supersymmetric gauged quantum mechanics and BPS states of four-dimensional $\mathcal{N}=2$ systems.  See\cite{Alim:2011kw}  for a systematic introduction and examples.
This connection motivates the analysis of the refined index $\Omega(y,\zeta)$ in a broad class of quantum-mechanical models. 

Fix a four-dimensional $\mathcal{N}=2$ system and a generic vacuum $v$ on its Coulomb branch.  At low energies, the physics is described by an abelian gauge theory with electromagnetic charge lattice $\Gamma$. The one-particle Hilbert space of the theory supports BPS states carrying charges $\gamma \in \Gamma$.  For each occupied charge the Hilbert space in that sector is a representation of $su(2)_{J}\times su(2)_{I},$ where $su(2)_{J}$ is group of spatial rotations and $su(2)_{I}$ is the $R$-symmetry of the four-dimensional theory.  This representation takes the general form
\begin{equation}
\left[\phantom{\int}\hspace{-.13in}(\mathbf{2},\mathbf{1})\oplus (\mathbf{1}, \mathbf{2})\right]\otimes \mathcal{H}_{\gamma}.
\end{equation}
We count BPS states by forming a protected spin character
\begin{equation}
\Omega(\gamma, y,v)_{4d}=\mathrm{Tr}_{\mathcal{H}_{\gamma}}y^{2J_{3}}(-y)^{2I_{3}}.
\end{equation}
$\Omega(\gamma, y,v)_{4d}$ receives contributions only from BPS states, and is stable under small variations in the vacuum $v$.  Under large changes in $v,$ $\Omega(\gamma, y,v)_{4d}$ may jump according to the wall-crossing formula \cite{Denef:2007vg, Kontsevich:2008fj, Joyce:2008pc}.

Next, let us describe an approach to the calculation of the protected spin characters $\Omega(\gamma, y,v)_{4d}$ utilizing supersymmetric quantum mechanics.  The basic physical paradigm of this method is to isolate a collection of elementary BPS states, and then to view the remaining BPS particles as non-relativistic composites of the elementary states.  Since the worldvolume theory of a BPS particle preserves four supercharges, the interactions governing the formation of non-relativistic bound states are controlled by $\mathcal{N}=4$ quantum mechanics.  Frequently this quantum mechanics is of the gauge theory type investigated in the previous section.

In a large class of models the relevant $\mathcal{N}=4$ quantum mechanics is a quiver model with unitary gauge groups and bifundamental matter.  In broad strokes, the dictionary between the two systems is as follows.  Each elementary constituent BPS state is represented by a node of the quiver giving a quantum mechanical gauge group.  The interactions between these nodes are encoded by the Dirac inner product of their electromagnetic charges and specify the number of arrows in the quiver.  In the quantum mechanics model, these are the chiral multiplets.  Finally, the central charges of the elementary BPS states map to the FI parameters $\zeta.$\footnote{If a superpotential is permitted by the topology of the quiver, then it must also be specified.  See e.g. \cite{Alim:2011ae} for a class of four-dimensional gauge theories where the relevant quiver superpotential may be fixed. } 

The main difficulty in applying the quantum-mechanical approach outlined above is to determine an explicit basis of elementary BPS states.  However in many four-dimensional theories, including for instance arbitrary gauge theories coupled to fundamental matter \cite{Alim:2011kw, Cecotti:2012gh}, such a basis may be identified and the BPS spectrum may be investigated.  When this is so we obtain a direct relationship between the four-dimensional protected spin character and the refined index of the associated gauged quantum mechanics:\footnote{The identification \eqref{protectedspin} suggests that in models for which the correspondence holds, all the ground states of the quantum mechanics are bosonic with vanishing $u(1)_{R}$ charge, and that the $su(2)_{I}$ charge acts trivially on the spectrum of BPS particles as in the ``no-exotics" conjecture of \cite{Gaiotto:2010be}. }
\begin{equation}
\Omega(\gamma, y,v)_{4d}=\Omega(y,\zeta), \label{protectedspin}
\end{equation}
where in the above we have the following explicit identification of parameters. 
  
\begin{itemize}
\item The $i$-th quiver gauge group is $U(n_{i})$ where the $n_{i}$ are determined by expanding the charge $\gamma$ as a sum of the charges of the elementary BPS states
\begin{equation}
\gamma =  \sum_{i}n_{i}\gamma_{i}. \label{chargesum}
\end{equation}

Since the matter content from the chiral multiplets consists of bifundamentals, the overall $U(1) \subset \prod_{i}U(n_{i})$ decouples and is treated as non-dynamical.  Alternatively, one may freely decouple any other convenient $U(1)$ without effecting the refined index.

\item The FI parameters are specified by the choice of vacuum $v$.  Each elementary BPS state has a central charge $\mathcal{Z}_{i}(v)$ which depends explicitly on $v$.  The central charge of $\gamma$ is then determined from \eqref{chargesum} by linearity
\begin{equation}
\mathcal{Z}(\gamma,v)=\sum_{i}n_{i}\mathcal{Z}_{i}(v) \equiv  |\mathcal{Z}(\gamma,v)|\exp(i\alpha), \hspace{.5in}\alpha \in \mathbb{R}.
\end{equation}
The FI parameter at the $i$-th node is then given by
\begin{equation}
\zeta_{i}=\Im\left(\phantom{\int}\hspace{-.16in}\exp(-i\alpha)n_{i}\mathcal{Z}_{i}(v)\right). \label{FIn2def}
\end{equation}
Observe that by construction, the sum of the FI parameters is zero.  This enables the decoupling of the overall $U(1)$ described above.
\end{itemize}

One interesting consequence of the identification \eqref{protectedspin} and the associated dictionary, is that the four-dimensional wall-crossing phenomenon maps to the discontinuity in the refined index $ \Omega(y,\zeta)$ under large changes in $\zeta$.  Because $\zeta$ enters our residue formula \eqref{residueform} as a definition of the contour, it follows that the four-dimensional wall-crossing formulas of \cite{Denef:2007vg, Kontsevich:2008fj, Joyce:2008pc} must be encoded in the variations of the residue integral as the contour is deformed.  This is similar the perspective on wall-crossing developed in systems with $\mathcal{N}=4$ supersymmetry in  \cite{Cheng:2007ch}.

\subsection{Toy Models}
In this section we describe simple examples of the relation between BPS particles and quiver quantum mechanics.  We study these models using the residue formula in \S \ref{examples}.

\subsubsection{Dyon Chains}
\label{chainsec}

A basic example illustrating the connection between four-dimensional BPS particles and ground states of supersymmetric gauged quantum mechanics are dyon chains.  These have been studied from the semiclassical soliton perspective in \cite{Stern:2000ie} and from the quiver quantum mechanics perspective in \cite{Denef:2002ru}.

The relevant four-dimensional system is $SU(M)$ super-Yang-Mills.  One is interested in investigating the bound states of a collection of $n+1\leq M$ distinct dyons.  We choose the electric and magnetic charges of the dyons as 
\begin{equation}
( e_{i}, m_{i})=( q_{i}\alpha_{i}, \alpha_{i}),
\end{equation}
where $\alpha_{i}$ denote simple roots of the $SU(M)$ algebra normalized such that
\begin{equation}
\alpha_{i}\cdot \alpha_{j}=\begin{cases} \phantom{-}2 & |i-j|=0, \\ -1 & |i-j|=1, \\ \phantom{-}0 & |i-j|>1, \end{cases}
\end{equation}
and $q_{i}$ are integers satisfying
\begin{equation}
q_{n+1}>q_{n} >\cdots > q_{3}>q_{2}>q_{1}.
\end{equation}

\begin{figure}[here!]
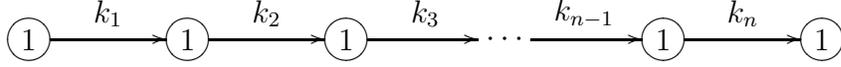

  \centerline{
{
\xy  0;<1pt,0pt>:<0pt,-1pt>::
(-300,0) *+{1}*\cir<8pt>{} ="1",
(-240,0) *+{1}*\cir<8pt>{} ="2",
(-180,0) *+{1}*\cir<8pt>{} ="3",
(-120,0) *+{\cdots} ="4",
(-60,0) *+{1}*\cir<8pt>{}="5",
(0,0) *+{1}*\cir<8pt>{}="6",
(-270, -10) *+{k_{1}} ="a",
(-210, -10) *+{k_{2}} ="b",
(-150, -10) *+{k_{3}} ="c",
(-90, -10) *+{k_{n-1}} ="d",
(-30, -10) *+{k_{n}} ="e",
\ar @{->} "1"; "2"
\ar @{->} "2"; "3"
\ar @{->} "3"; "4"
\ar @{->} "4"; "5"
\ar @{->} "5"; "6"
\endxy}}
  \caption{The general abelian linear quiver which governs the bounds states of the specified dyons. The integers at the nodes denote the ranks of the associated gauge groups, while $k_{i}$ are the number of bifundamentals (arrows).}
  \label{fig:chain}
\end{figure}

If we denote by $k_{i}$ the quantity $q_{i+1}-q_{i}$, then the symplectic products of the dyon charges are
\begin{equation}
( e_{i}, m_{i}) \cdot ( e_{j}, m_{j})= \begin{cases}  + k_{i} & j=i + 1, \\ - k_{i-1} & j=i - 1, \\    \phantom{\pm} 0 & j \neq i \pm 1.\end{cases}
\end{equation}
The quiver governing the bound states of these dyons is then a linear chain illustrated in Figure \ref{fig:chain}.

The spectrum of bound states depends on the FI parameters $\zeta_{i}$ at the $i$-th node.  When these are such that 

\begin{equation}
\zeta_{n+1}>0, \hspace{.5in}\zeta_{n+1}+\zeta_{n}>0,\hspace{.5in}\cdots \hspace{.5in}\zeta_{n+1}+\zeta_{n}+\cdots+\zeta_{2} >0,
\end{equation} 
there is a non-trivial classical Higgs branch, $\mathcal{M},$ of supersymmetric vacua in the quiver.  By explicitly solving the $F$ and $D$ term equations of \S \ref{cohomology}, one finds that the Higgs branch is a product of projective spaces
\begin{equation}
\mathcal{M}= \prod_{i=1}^{n}\mathbb{P}^{k_{i}-1}
\end{equation}
quantizing this space as in \eqref{cohomindex} we find that the index is 
\begin{equation}
\Omega(y,\zeta) =\prod_{i=1}^n \left( y^{-k_i+1}\sum_{j=0}^{k_i-1} y^{2j}\right), \label{linabanswer}
\end{equation}
which reproduces the answers obtained by quantizing monopole moduli spaces \cite{Stern:2000ie}.  

We obtain this result using the residue formula \eqref{residueform} in \S \ref{linab}.

\subsubsection{Electron Halos}
\label{halosec}

\begin{figure}[here!]
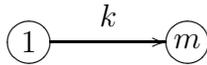

  \centerline{
{
\xy  0;<1pt,0pt>:<0pt,-1pt>::
(-300,0) *+{1}*\cir<8pt>{} ="1",
(-240,0) *+{m}*\cir<8pt>{} ="2",
(-270, -10) *+{k} ="a",
\ar @{->} "1"; "2"
\endxy}}
  \caption{The quiver relevant for studying the bound states of a monopole and a cloud of electrons.  The integers at the nodes denote the ranks of the associated gauge groups, while $k$ is the number of bifundamentals (arrows). }
  \label{fig:halo}
\end{figure}

Another class of interesting examples arises from studying the bound states of $m$ identical electrons and a single monopole of magnetic charge $k$.  In this case, the relevant quiver is shown in Figure \ref{fig:halo}.

Let $\zeta$ indicate the FI parameter at the second node and assume $\zeta >0$.  By solving the equations of \S \ref{cohomology}, we determine that the moduli space is the Grassmannian $Gr(m, k)$ of complex $m$-planes in a $k$-dimensional space.\footnote{The moduli space is empty if $m>k$.}  Extracting the refined index from cohomology as in \eqref{cohomindex}, we find that 
\begin{equation}
\Omega(y, \zeta) = {y^{m(m-k)}\prod_{i=1}^k (1-y^{2i}) \over \prod_{i=1}^m (1-y^{2i})\prod_{i=1}^{k-m} (1-y^{2i}) }. \label{grassmannanswer}
\end{equation}
We reproduce this result using the residue formula \eqref{residueform} in \S\ref{haloresiduecomp}.

\section{Examples}
\label{examples}
In this section we explore various examples of the residue formula \eqref{residueform} for the refined index $\Omega(y,\zeta)$.  The cases we consider illustrate several interesting features of the index: wall-crossing, non-Abelian gauge groups, and superpotentials.  

To achieve maximal overlap with the applications discussed in \S\ref{BPSStates}, we consider quantum-mechanical quiver gauge theories with unitary gauge groups.  In such examples a single $U(1)$ factor of the gauge group decouples.  One may choose this $U(1)$ to simplify the resulting quantum mechanics. Correspondingly, we demand that the sum of the FI parameters vanishes as in \eqref{FIn2def}.

\subsection{Linear Abelian Quivers: Dyon Chains}
\label{linab}

We begin with the example of linear abelian quivers.  As described in \S \ref{chainsec}, these quivers compute the bound states of chains of distinct dyons. We aim to reproduce the result \eqref{linabanswer} using the residue formula \eqref{residueform}.

\subsubsection{Two Nodes}

\begin{figure}[here!]
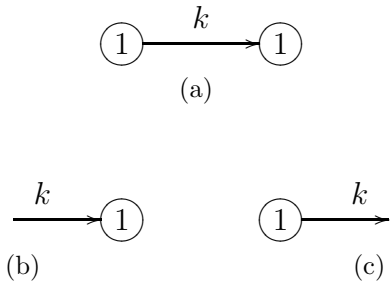

  \centering
\subfloat[]{\label{fig:L2node}
\xy  0;<1pt,0pt>:<0pt,-1pt>::
(-300,0) *+{1}*\cir<8pt>{} ="1",
(-240,0) *+{1}*\cir<8pt>{} ="2",
(-270, -10) *+{k} ="a",
\ar @{->} "1"; "2"
\endxy}
\\
\vspace{.2in}
\subfloat[]{\label{fig:L2nodedec1}
\xy  0;<1pt,0pt>:<0pt,-1pt>::
(-300,0) *+{\phantom{A+A}}="1",
(-240,0) *+{1}*\cir<8pt>{} ="2",
(-270, -10) *+{k} ="a",
\ar @{->} "1"; "2"
\endxy}
\hspace{.5in}
\subfloat[]{\label{fig:L2nodedec2}
\xy  0;<1pt,0pt>:<0pt,-1pt>::
(-300,0) *+{1}*\cir<8pt>{} ="1",
(-240,0) *+{\phantom{A+A}} ="2",
(-270, -10) *+{k} ="a",
\ar @{->} "1"; "2"
\endxy}
  \caption{The two-node linear quiver.  The integers at the nodes denote the ranks of the associated gauge groups, while $k$ is the number of bifundamentals (arrows).  In (b) and (c), the two ways of decoupling a $U(1).$}\label{2nodequiver}
\end{figure}

We start with the abelian two-node quiver with $k$ bifundamental chiral multiplets between the two nodes. We can decouple a $U(1)$ in two different ways as shown in Figure \ref{fig:L2node}. We decouple the first node as in Figure \ref{fig:L2nodedec1}. The other alternative clearly yields the same answer.

In this case the one-loop determinant is 
\begin{align}
Z_{1-\text{loop}} (z,u) =  - {\pi\over \sin (\pi z)} \left[ {\sin(\pi u-\pi z)\over \sin (\pi u) }\right]^kdu.
\end{align}
On $\frak{M}$, there is a hyperplane $H$ (in this case, point) where $Z_{1-\text{loop}}$ has a pole:
\begin{align}
H:~u=0.
\end{align}
The corresponding charge covector $Q$ is just 1. Let $\zeta_2$ be the FI parameter of the second node. The Jeffrey-Kirwan residue operation satisfies
\begin{align}
 \underset{u=0}{\text{JK-Res}}\left(
\{1\} ,  \zeta_2   \right) {du\over  u} = 
\begin{cases}
1,&\text{if}~ \zeta_2>0,\\
0,&\text{if}~\zeta_2<0.
\end{cases}
\end{align}
In the $\zeta_2>0$ case, we can therefore write the Jeffrey-Kirwan residue as the usual contour integral around $H=\{\,u=0\,\}$:
\begin{align}
 \underset{u=0}{\text{JK-Res}}\left(
\{1\} ,  \zeta_2   \right) {du\over  u}
= {1\over 2\pi i} \oint_{u=0} {du\over u},~~~~\text{if}~\zeta_2>0.
\end{align} 

The index is then given by
\begin{align}
\Omega(y,\zeta_2)&=- {\pi \over \sin (\pi z)}
 \underset{u=0}{\text{JK-Res}}\left(
\{1\} , \zeta_2   \right)
\left[
{\sin(\pi u - \pi z)\over \sin(\pi u)}
\right]^k
du\nonumber\\
&
= \begin{cases}
- {\pi \over \sin (\pi z)}
\oint_{u=0} {du \over 2\pi i}
\left[
{\sin(\pi u - \pi z)\over \sin(\pi u)}
\right]^k
~~&\text{if}~\zeta_2>0,\nonumber\\
0~~&\text{if}~\zeta_2<0.
\end{cases}\nonumber\\
\label{linonenodesans}
& =\begin{cases}
 y^{-k+1}\sum_{j=0}^{k-1} y^{2j}~~&\text{if}~\zeta_2>0,\\
 0~~&\text{if}~\zeta_2<0.
 \end{cases}
\end{align}

\subsubsection{Three Nodes}

\begin{figure}[here!]
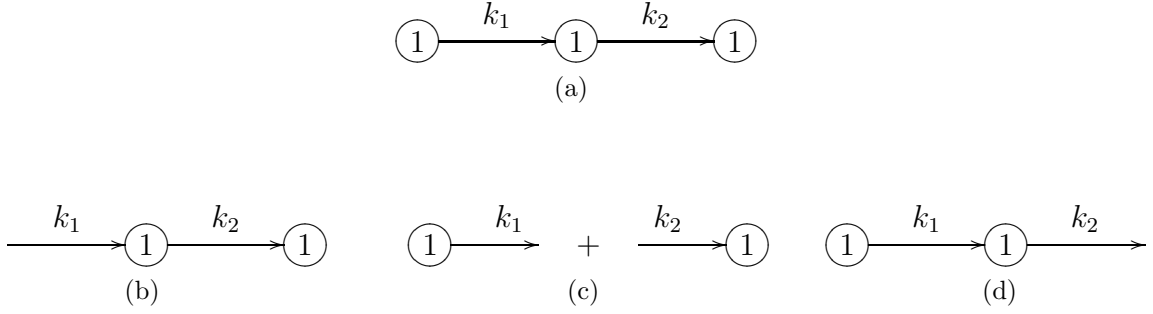

  \centering
\subfloat[]{\label{fig:L3node}
\xy  0;<1pt,0pt>:<0pt,-1pt>::
(-300,0) *+{1}*\cir<8pt>{} ="1",
(-240,0) *+{1}*\cir<8pt>{} ="2",
(-180,0) *+{1}*\cir<8pt>{} ="3",
(-270, -10) *+{k_{1}} ="a",
(-210, -10) *+{k_{2}} ="b",
\ar @{->} "1"; "2"
\ar @{->} "2"; "3"
\endxy}
\\
\vspace{.33in}
\subfloat[]{\label{fig:L3nodedec1}
\xy  0;<1pt,0pt>:<0pt,-1pt>::
(-300,0) *+{\phantom{A}} ="1",
(-240,0) *+{1}*\cir<8pt>{} ="2",
(-180,0) *+{1}*\cir<8pt>{} ="3",
(-270, -10) *+{k_{1}} ="a",
(-210, -10) *+{k_{2}} ="b",
\ar @{->} "1"; "2"
\ar @{->} "2"; "3"
\endxy}
\hspace{.33in}
\subfloat[]{\label{fig:L3nodedec2}
\xy  0;<1pt,0pt>:<0pt,-1pt>::
(-300,0) *+{1}*\cir<8pt>{} ="1",
(-240,0) *+{\phantom{A}+\phantom{A}} ="2",
(-180,0) *+{1}*\cir<8pt>{} ="3",
(-270, -10) *+{k_{1}} ="a",
(-210, -10) *+{k_{2}} ="b",
\ar @{->} "1"; "2"
\ar @{->} "2"; "3"
\endxy}
\hspace{.2in}
\subfloat[]{\label{fig:L3nodedec3}
\xy  0;<1pt,0pt>:<0pt,-1pt>::
(-300,0) *+{1}*\cir<8pt>{} ="1",
(-240,0) *+{1}*\cir<8pt>{} ="2",
(-180,0) *+{\phantom{A}} ="3",
(-270, -10) *+{k_{1}} ="a",
(-210, -10) *+{k_{2}} ="b",
\ar @{->} "1"; "2"
\ar @{->} "2"; "3"
\endxy}
  \caption{The three-node linear quiver.  The integers at the nodes denote the ranks of the associated gauge groups, while $k_{i}$ are the number of bifundamentals (arrows).  In (b), (c), and (d), the three ways of decoupling a $U(1).$  In (b), the quiver has become disconnected and the model factorizes.}
\end{figure}

Let us now move on to the three-node linear quiver with $k_i$ bifundamental chiral multiplets between the $i$-th and the $(i+1)$-th nodes. There are three distinct ways to decouple a $U(1)$ from the quiver as shown in Figure \ref{fig:L3node}. For purposes of illustration we will show explicitly that all three choices yield the same answer. 

The easiest choice is to decouple the second node as in Figure \ref{fig:L3nodedec2}, so that the quiver becomes two decoupled one-node quivers. The index is immediately given by the product of the answers \eqref{linonenodesans} for the one-node quivers:
\begin{align}\label{second}
\Omega(y,\zeta) = \begin{cases}
\left(  y^{-k_1+1}\sum_{i=0}^{k_1-1} y^{2i}\right)\left( y^{-k_2+1}\sum_{j=0}^{k_2-1} y^{2j}\right), &\text{if}~\zeta_1<0,~\zeta_3>0,\\
0,&\text{otherwise}.
\end{cases}
\end{align}

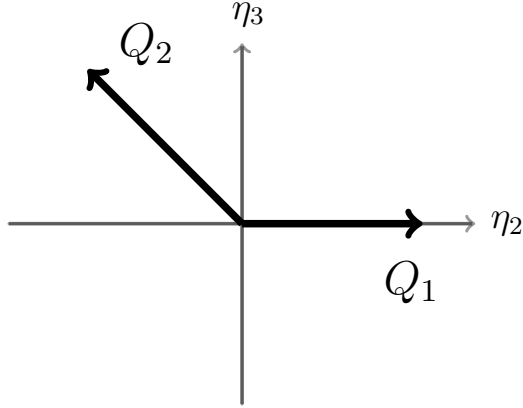
\begin{figure}
\begin{center}
\begin{tikzpicture}[scale=2]
\coordinate (O) at (0,0);
\coordinate (A2)  at (135:1.45);
\draw[line width=1mm,black,->] (O)--(A2);
\draw[line width=1mm,black,->] (O)--(0:1.2);
\draw[line width=0.5mm,black,opacity=.4,->] (0:-1.55)--(0:1.55);
\draw[line width=0.5mm,black,opacity=.4,->] (-90:1.2)--(90:1.2);
\node at (118:1.35) {\color{black} \scalebox{1.5}{$Q_2$}};
\node at (-23:1) {\color{black} \scalebox{1.5}{~~~~$Q_1$}};
\node at (0:1.6) {\color{black} \scalebox{1.2}{~~~~$\eta_2$}};
\node at (95:1.4) {\color{black} \scalebox{1.2}{~~~~$\eta_3$}};
\end{tikzpicture}
\caption{Three-node quiver with the first node decoupled. The figure shows the charge covectors $Q_i$ in $\frak{h}^*= (\frak{u}(1)^2)^*\cong \mathbb{R}^2$.}\label{3node}
\end{center}
\end{figure}

Alternatively, we can decouple the first node as in Figure \ref{fig:L3nodedec1}. The one-loop determinant is
\begin{align}
Z_{1-\text{loop}}(z,u)=\left[
{\sin(\pi u_2 - \pi z)\over \sin(\pi u_2)}
\right]^{k_1}
\left[
{\sin(-\pi u_2+\pi u_3 - \pi z)\over \sin(-\pi u_2+\pi u_3)}
\right]^{k_2}
du_2\wedge du_3.
\end{align}
There are two  hyperplanes on the complex two-dimensional space $\frak{M}$ where $Z_{1-\text{loop}}$ has poles:
\begin{align}
\begin{split}
&H_1:~ u_2=0,\\
&H_2:~-u_2+u_3=0.
\end{split}
\end{align}
The corresponding charge covectors $Q_i$ that define $H_i$ in $\frak{M}$ are
\begin{align}
\begin{split}
&Q_1 = ( 1,0),\\
&Q_2=(-1,1),
\end{split}
\end{align}
as shown in Figure \ref{3node}. The intersection $H_1\cap H_2= \{ u=0\}$ is the point $u_*$ at which we evaluate the residue. Since this theory is abelian, $\frak{g}_{u(1)}^* = \frak{h}^*$ and we can take $\eta= \zeta$ to be on any point on the $\frak{h}^*$ plane in Figure \ref{3node}. 

From the definition of the Jeffrey-Kirwan residue operation, we have
\begin{align}
 &\underset{u=0}{\text{JK-Res}}\left(
\{Q_1,Q_2\} , \zeta   \right)
{du_2 \wedge du_3\over u_2(-u_2+u_3)} 
= 
\begin{cases}
1 ,&\text{if}~ \zeta \in \text{Cone}(Q_1,Q_2),\\
0,&\text{otherwise}.
\end{cases}\end{align}
If $ \zeta \in \text{Cone}(Q_1,Q_2)$, we can then write the Jeffrey-Kirwan residue as
\begin{align}
 \underset{u=0}{\text{JK-Res}}\left(
\{Q_1,Q_2\} , \zeta   \right)
{du_2 \wedge du_3\over u_2(-u_2+u_3)} 
= 
\left( {1\over 2\pi i}\right)^2    \oint_{u_2=0}\oint_{u_3=u_2} {du_2du_3 \over u_2(-u_2+u_3)}.
\end{align}

The index in the chamber $\zeta\in \text{Cone} (Q_1,Q_2)$ is
\begin{align}
\Omega(y,\zeta)&
=\left[- {\pi \over \sin (\pi z)}\right]^2
\oint_{u_2=0}{du_2\over 2\pi i} \oint_{u_3= u_2} {du_3\over 2\pi i} \, 
\left[
{\sin(\pi u_2 - \pi z)\over \sin(\pi u_2)}
\right]^{k_1}
\left[
{\sin(-\pi u_2+\pi u_3 - \pi z)\over \sin(-\pi u_2+\pi u_3)}
\right]^{k_2}
\nonumber\\
&=\left(  y^{-k_1+1}\sum_{i=0}^{k_1-1} y^{2i}\right)\left( y^{-k_2+1}\sum_{j=0}^{k_2-1} y^{2j}\right).
\end{align}
The condition $\zeta \in \text{Cone}(Q_1,Q_2)$ for nonzero index in components is
\begin{align}
\zeta_3>0,~~\zeta_2+\zeta_3>0,
\end{align}
which is the same as the answer obtained by decoupling the second node in \eqref{second}. Note that we have used $\zeta_1+\zeta_2+\zeta_3=0$. Similarly one can show that the index obtained by decoupling the $U(1)$ as in Figure \ref{fig:L3nodedec3} is the same as above.

\subsubsection{General Linear Abelian Quiver}

\begin{figure}[here!]
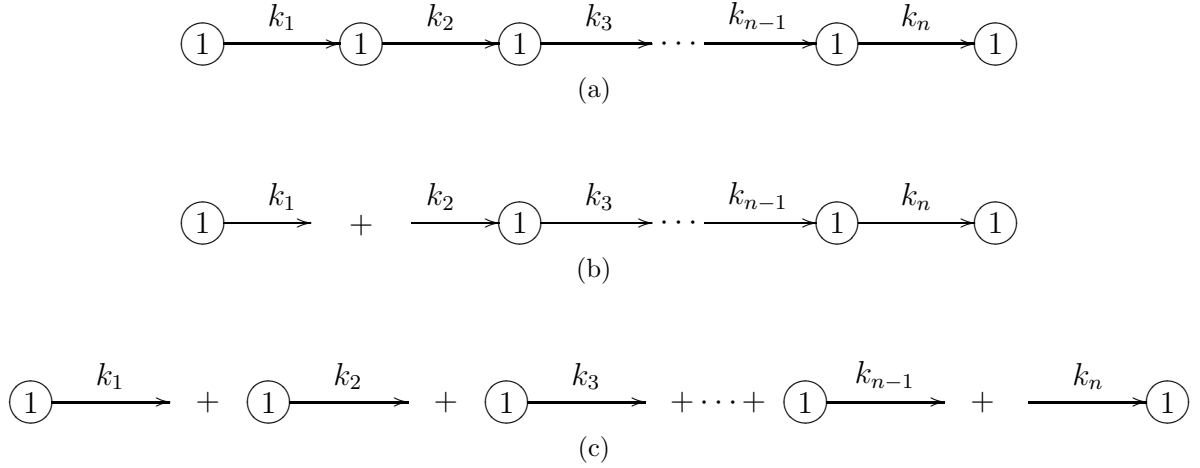

  \centering
\subfloat[]{\label{fig:Lgen}
\xy  0;<1pt,0pt>:<0pt,-1pt>::
(-300,0) *+{1}*\cir<8pt>{} ="1",
(-240,0) *+{1}*\cir<8pt>{} ="2",
(-180,0) *+{1}*\cir<8pt>{} ="3",
(-120,0) *+{\cdots} ="4",
(-60,0) *+{1}*\cir<8pt>{}="5",
(0,0) *+{1}*\cir<8pt>{}="6",
(-270, -10) *+{k_{1}} ="a",
(-210, -10) *+{k_{2}} ="b",
(-150, -10) *+{k_{3}} ="c",
(-90, -10) *+{k_{n-1}} ="d",
(-30, -10) *+{k_{n}} ="e",
\ar @{->} "1"; "2"
\ar @{->} "2"; "3"
\ar @{->} "3"; "4"
\ar @{->} "4"; "5"
\ar @{->} "5"; "6"
\endxy}
\\
\vspace{.2in}
\subfloat[]{\label{fig:Lgendec1}
\xy  0;<1pt,0pt>:<0pt,-1pt>::
(-300,0) *+{1}*\cir<8pt>{} ="1",
(-240,0) *+{\phantom{A}+\phantom{A}}="2",
(-180,0) *+{1}*\cir<8pt>{} ="3",
(-120,0) *+{\cdots} ="4",
(-60,0) *+{1}*\cir<8pt>{}="5",
(0,0) *+{1}*\cir<8pt>{}="6",
(-270, -10) *+{k_{1}} ="a",
(-210, -10) *+{k_{2}} ="b",
(-150, -10) *+{k_{3}} ="c",
(-90, -10) *+{k_{n-1}} ="d",
(-30, -10) *+{k_{n}} ="e",
\ar @{->} "1"; "2"
\ar @{->} "2"; "3"
\ar @{->} "3"; "4"
\ar @{->} "4"; "5"
\ar @{->} "5"; "6"
\endxy}
\\
\vspace{.2in}
\subfloat[]{\label{fig:Lgendec2}
\xy  0;<1pt,0pt>:<0pt,-1pt>::
(-300,0) *+{1}*\cir<8pt>{} ="1",
(-240,0) *+{\phantom{A}} ="2",
(-233,0) *+{+},
(-270, -10) *+{k_{1}}, 
(-210,0) *+{1}*\cir<8pt>{} ="3",
(-150,0) *+{\phantom{A}} ="4",
(-143,0) *+{+},
(-180, -10) *+{k_{2}}, 
(-120,0) *+{1}*\cir<8pt>{} ="5",
(-60,0) *+{\phantom{A}} ="6",
(-40,0) *+{+ \cdots +},
(-90, -10) *+{k_{3}}, 
(-7,0) *+{1}*\cir<8pt>{} ="7",
(53,0) *+{\phantom{A}} ="8",
(60,0) *+{+},
(23, -10) *+{k_{n-1}}, 
(70,0) *+{\phantom{A}} ="9",
(130,0) *+{1}*\cir<8pt>{} ="10",
(100, -10) *+{k_{n}}, 
\ar @{->} "1"; "2"
\ar @{->} "3"; "4"
\ar @{->} "5"; "6"
\ar @{->} "7"; "8"
\ar @{->} "9"; "10"
\endxy}
  \caption{The general abelian linear quiver. The integers at the nodes denote the ranks of the associated gauge groups, while $k_{i}$ are the number of bifundamentals (arrows).  In (b), a convenient choice of decoupled $U(1)$.  In (c), the model is reduced to a product.}
\end{figure}

Finally, we consider the general abelian linear quiver with $n+1$ nodes and $k_i$ bifundamental chiral multiplets between the $i$-th and the $(i+1)$-th nodes. 

In the abelian three-node quiver case $(n=2)$, we have shown that $\Omega(y,\zeta)$ is the product of the index of the one-node quiver with $k_1$ chiral multiplets, and the index for the one-node quiver with $k_2$ chiral multiplets. 

Now assume that for the $n$-node quiver $\Omega(y,\zeta)$ is similarly given by the product of that for $n-1$ one-node quivers with $k_i$ chiral multiplets for the $i$-th decoupled node. Then for the linear quiver with $n+1$ nodes, we can decouple the second node as shown in Figure \ref{fig:Lgendec1} and the quiver becomes the product of a one-node quiver with a $n$-node quiver. Inductively, we have shown that the index for the $(n+1)$-node quiver is the same as the product of indices for $n$ one-node quivers as shown in Figure \ref{fig:Lgendec2}.

Thus, the index of the general abelian linear quiver is:
\begin{align}
\Omega(y,\zeta) =\prod_{i=1}^n \left( y^{-k_i+1}\sum_{j=0}^{k_i-1} y^{2j}\right),
\end{align}
if the FI parameters $\zeta_i$ satisfy the following conditions
\begin{equation}
\zeta_{n+1}>0, \hspace{.5in}\zeta_{n+1}+\zeta_{n}>0,\hspace{.5in}\cdots \hspace{.5in}\zeta_{n+1}+\zeta_{n}+\cdots+\zeta_{2} >0,
\end{equation} 
as can be easily seen from Figure \ref{fig:Lgendec1}.

This is exactly the expected result \eqref{linabanswer}.

\subsection{Non-Abelian Phenomena: Electron Halos}
\label{haloresiduecomp}

In this section we consider an example with non-abelian quiver gauge group.  As described in \S\ref{halosec} this example computes the bound states of a single monopole and $m$ identical electrons.  Our goal is to reproduce the result \eqref{grassmannanswer} using the residue formula \eqref{residueform}.

\begin{figure}[here!]
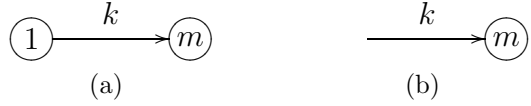

  \centering
\subfloat[]{\label{fig:U(m)a}
\xy  0;<1pt,0pt>:<0pt,-1pt>::
(-300,0) *+{1}*\cir<8pt>{} ="1",
(-240,0) *+{m}*\cir<8pt>{} ="2",
(-270, -10) *+{k} ="a",
\ar @{->} "1"; "2"
\endxy}
\hspace{.5in}
\subfloat[]{\label{fig:U(m)b}
\xy  0;<1pt,0pt>:<0pt,-1pt>::
(-300,0) *+{\phantom{A}}="1",
(-240,0) *+{m}*\cir<8pt>{} ="2",
(-270, -10) *+{k} ="a",
\ar @{->} "1"; "2"
\endxy}
  \caption{The two-node linear quiver with a nonabelian gauge group.  The integers at the nodes denote the ranks of the associated gauge groups, while $k$ is the number of bifundamentals (arrows).  In (b), a $U(1)$ is decoupled leaving a $U(m)$ gauge theory with $k$ fundamental chiral multiplets with +1 charge under the $U(1)$ of $U(m)$.}
\end{figure}

Consider the quiver in Figure \ref{fig:U(m)a}. We decouple the $U(1)$ node to compute the index as in Figure \ref{fig:U(m)b}. One can alternatively decouple the central $U(1)$ of $U(m)$ and obtain the same answer. 

The one-loop determinant for a $U(m)$ vector multiplet with $k$ chiral multiplets in the representation \scalebox{.5}{\,\ydiagram{1}\,}$_1$ is
\begin{align}
Z_{1-\text{loop}}={1\over m!}\left[ - {\pi \over \sin (\pi z)}\right]^m 
\prod_{\substack{b,c=1,\\ b\neq c}}^m {\sin(\pi u_b  - \pi u_c )\over \sin(\pi u_b -\pi u_c -\pi z) } 
\prod_{ a=1}^m \left[{\sin(\pi u_a-\pi z)\over \sin(\pi u_a)}\right]^k \,
du_1\wedge\cdots \wedge du_m.
\end{align}
On the complex $m$-dimensional space $\frak{M}$, there are hyperplanes $H_{ab}$ and $H_c$, with $a,b,c=1,\cdots,m$ and $a\neq b$, where $Z_{1-\text{loop}}$ has poles:
\begin{align}
\begin{split}
&\text{vector}:~H_{ab}:~u_a - u_b -z =0,~~~a\neq b,\\
&\text{chiral}:~H_c :~ u_c=0.
\end{split}
\end{align}

For the index formula, we always pick the covector $\eta$ in the definition of the Jeffrey-Kirwan residue operation \eqref{JK} to be in the $u(1)$ part of the dual Cartan subalgebra $\frak{g}_{u(1)}^*$ as in \eqref{etazeta}. In the current example, this implies that $\eta$ lies on a real one-dimensional line on the real $m$-dimensional space $\frak{h}^*$:
\begin{align}
\eta= \zeta ( 1,1,\cdots,1)\in  \frak{h}^*\cong \mathbb{R}^m,
\end{align}
where $\zeta$ is the FI parameter for $U(m)$.

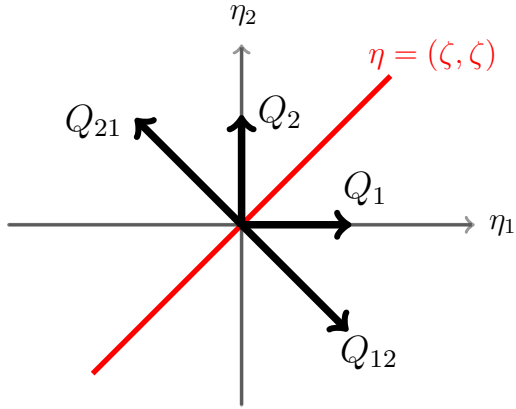
\begin{figure}[here!]
\begin{center}
\begin{tikzpicture}[scale=2]
\coordinate (O) at (0,0);
\coordinate (A2)  at (135:1);
\coordinate (A1) at (-45:1);
\draw[line width=0.7mm,red,opacity=1] (225:1.4)--(45:1.4);
\draw[line width=1mm,black,->] (O)--(A2);
\draw[line width=1mm,black,->] (O)--(A1);
\draw[line width=1mm,black,->] (O)--(0:.73);
\draw[line width=1mm,black,->] (O)--(90:.73);
\draw[line width=0.5mm,black,opacity=.4,->] (0:-1.55)--(0:1.55);
\draw[line width=0.5mm,black,opacity=.4,->] (-90:1.2)--(90:1.2);
\node at (-45:1.2) {\color{black} \scalebox{1.2}{$Q_{12}$}};
\node at (145:1.2) {\color{black} \scalebox{1.2}{$Q_{21}$}};
\node at (20:.7) {\color{black} \scalebox{1.2}{~~~~$Q_1$}};
\node at (83:.75) {\color{black} \scalebox{1.2}{~~~~$Q_2$}};
\node at (0:1.6) {\color{black} \scalebox{1}{~~~~$\eta_1$}};
\node at (95:1.4) {\color{black} \scalebox{1}{~~~~$\eta_2$}};
\node at (45:1.6) {\color{red} \scalebox{1}{~~~~$\eta = (\zeta,\zeta)$}};
\end{tikzpicture}
\end{center}
\caption{A $U(2)$ vector multiplet with $k$ chiral multiplets in the fundamental representation with +1 charge under the $U(1)$ of $U(2)$. The figure shows the charge covectors $Q_{ab}$ and $Q_a$ on $\frak{h}^* \cong \mathbb{R}^2$. In the index formula, we choose $\eta$ to be on $\frak{g}_{u(1)}^*$, which is the red line in the figure. As a result, we never need to consider the chamber Cone$(Q_1, Q_{12})$ nor Cone$(Q_2,Q_{21})$.}\label{U(2)}
\end{figure}

For a given $\zeta$, the index can potentially receive contribution from various intersections of $H_{ab}$ and $H_a$. For example, in the $U(2)$ case shown in Figure \ref{U(2)}, if we choose $\zeta>0$, the Jeffrey-Kirwan residue operation receives contributions from $H_1\cap H_2$, $H_{12}\cap H_2$, and $H_{21}\cap H_1$, while it gives zero for $\zeta<0$. However, the contributions from $H_{12}\cap H_2$ and $H_{21}\cap H_1$ can be shown to be zero by a direct computation. 

For general $m$ in the chamber $\zeta>0$, we therefore conjecture that the index only receives contribution from the intersection $H_1\cap H_2\cap\cdots\cap H_m$.

With this assumption, the index can then be computed to be
\begin{align}\label{U(m)index1}
\Omega(y,\zeta)=&{1\over m!}\left[ - {\pi \over \sin (\pi z)}\right]^m 
\prod_{a=1}^m\left[  \oint_{u_a=0} {du_a \over 2\pi i} \right] \,
\prod_{\substack{b,c=1,\\ b\neq c}}^m {\sin(\pi u_b  - \pi u_c )\over \sin(\pi u_b -\pi u_c -\pi z) } 
\prod_{ d=1}^m \left[{\sin(\pi u_d-\pi z)\over \sin(\pi u_d)}\right]^k.
\end{align}
On the other hand, from the result \eqref{grassmannanswer} we know that the index is given by
\begin{align}\label{U(m)index2}
\Omega(y,\zeta)
&=\begin{cases}
{y^{m(m-k)}\prod_{i=1}^k (1-y^{2i}) \over \prod_{i=1}^m (1-y^{2i})\prod_{i=1}^{k-m} (1-y^{2i}) }~~&\text{if}~ m\le k,\\
0~~&\text{if}~m>k,
\end{cases}
\end{align}
when $\zeta>0$. 

We have checked by direct calculation that the two expressions \eqref{U(m)index1} and \eqref{U(m)index2} agree for a wide range of $m$ and $k$. This provides further evidence that index only receives contribution from the intersection $H_1\cap H_2\cap\cdots\cap H_m$ and yields an elegant combinatorial identity for the residue integral \eqref{U(m)index1}.

\subsection{Non-Trivial Superpotentials}
\label{potentialsec}
In the examples of \S\ref{linab} and \S\ref{haloresiduecomp} the quivers do not admit non-trivial superpotentials and hence the refined index is not sensitive to the choice of $u(1)_{R}$ charge assignments for the chiral multiplets.  In this section we generalize to examples where the superpotential plays an important role.  We find that as long as the superpotential satisfies the properties described in \S \ref{lagrangians} the index formula \eqref{residueform} still accurately computes the refined index.  The examples we explore fall into the class of quivers analyzed from a representation theory perspective in \cite{Denef:2007vg, Lee:2012sc, Lee:2012naa}.

\subsubsection{The $XYZ$ Model}

\begin{figure}[here!]
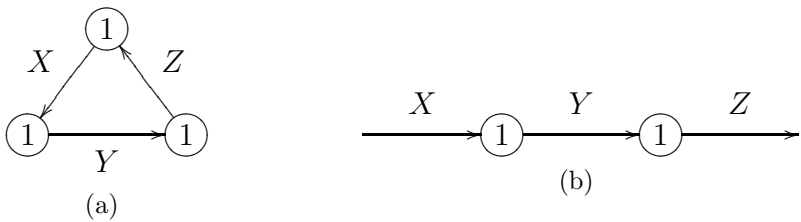

  \centering
\subfloat[]{\label{fig:triangle1}
\xy  0;<1pt,0pt>:<0pt,-1pt>::
(-300,0) *+{1}*\cir<8pt>{} ="1",
(-240,0) *+{1}*\cir<8pt>{} ="2",
(-270,-40) *+{1}*\cir<8pt>{} ="3",
(-270, 10) *+{Y} ="b",
(-295, -28) *+{X} ="a",
(-245, -28) *+{Z} ="c",
\ar @{->} "1"; "2"
\ar @{->} "3"; "1"
\ar @{->} "2"; "3"
\endxy}
\hspace{.5in}
\subfloat[]{\label{fig:triangle2}
\xy  0;<1pt,0pt>:<0pt,-1pt>::
(-270, -12) *+{X} ="x",
(-210, -12) *+{Y} ="y",
(-150, -12) *+{Z} ="z",
(-300,0) *+{\phantom{A}}="1",
(-240,0) *+{1}*\cir<8pt>{} ="2",
(-180,0) *+{1}*\cir<8pt>{} ="3",
(-120,0) *+{\phantom{B}}="4",
\ar @{->} "1"; "2",
\ar @{->} "2"; "3",
\ar @{->} "3"; "4",
\endxy}
  \caption{The $XYZ$ model. The integers at the nodes denote the ranks of the associated gauge groups, while $X, Y, Z$ label the fields. There is one arrow between each pair of nodes. In (b), a choice of $U(1)$ decoupling.}
\end{figure}

Consider a triangle quiver shown in Figure \ref{fig:triangle1} with three $U(1)$ vector multiplets and three chiral multiplets $X,Y,Z$. We decouple the $U(1)$ node where $X$ and $Z$ meet as in Figure \ref{fig:triangle2}.

We will assume the $R$-charges for the three chiral multiplets $X,Y,Z$ to be $R_X, \, R_Y,\, R_Z$, respectively. Let $k$ be a positive integer such that
\begin{align}
{2\over k} = R_X +R_Y+R_Z.
\end{align}
Given a $k$, we can allow for the following superpotential in the quantum mechanics:
\begin{align}
\mathcal{W} = (XYZ)^k.
\end{align}

For $k=1$ we have the generic cubic superpotential $\mathcal{W}=XYZ$.  There are no supersymmetric ground states, so the expected answer for $\Omega(y,\zeta)$ is zero.  In this case we will see that the residue formula \eqref{residueform} accurately computes the index.

When $k>1,$ the superpotential does not satisfy our hypotheses.  A direct calculation in the chamber where $\zeta_{1}>0$ and  $\zeta_{2}>0$ shows that the expected index from quantizing the classical moduli space is one.  We will see that the residue formula does not produce this answer.

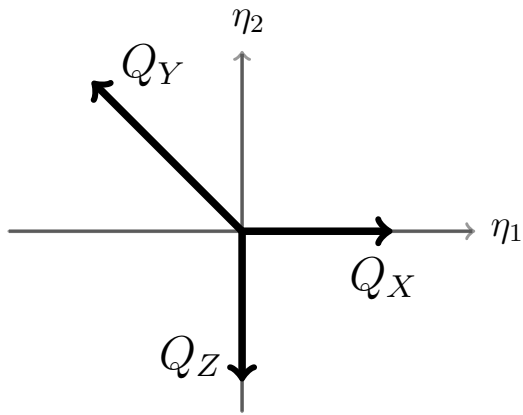
\begin{figure}[here!]
\begin{center}
\begin{tikzpicture}[scale=2]
\coordinate (O) at (0,0);
\coordinate (A2)  at (135:1.41);
\draw[line width=1mm,black,->] (O)--(A2);
\draw[line width=1mm,black,->] (O)--(0:1);
\draw[line width=0.5mm,black,opacity=.4,->] (0:-1.55)--(0:1.55);
\draw[line width=0.5mm,black,opacity=.4,->] (-90:1.2)--(90:1.2);
\draw[line width=1mm,black,opacity=1,->] (0:0)--(270:1);
\node at (118:1.25) {\color{black} \scalebox{1.5}{$Q_Y$}};
\node at (-23:.8) {\color{black} \scalebox{1.5}{~~~~$Q_X$}};
\node at (237:1) {\color{black} \scalebox{1.5}{~~~~$Q_Z$}};
\node at (0:1.6) {\color{black} \scalebox{1.2}{~~~~$\eta_1$}};
\node at (95:1.4) {\color{black} \scalebox{1.2}{~~~~$\eta_2$}};
\end{tikzpicture}
\caption{The $XYZ$ model with one node removed. The figure shows the charge covectors $Q_X,\, Q_Y,\, Q_Z$ on $\frak{h}^*\cong\mathbb{R}^2$.  }\label{triangle}
\end{center}
\end{figure}

The one-loop determinant is
\begin{align}
Z_{1-\text{loop}} &= \left[ - {\pi \over \sin(\pi z)}\right]^2
\left[ {\sin\left(\pi u_1+\pi  \left( {R_X\over2} -1\right)z\right)\over \sin(\pi u_1+ \pi {R_X\over 2}z)}\right]
\left[ {\sin\left(-\pi u_1+\pi u_2+\pi  \left( {R_Y\over2} -1\right)z\right)\over \sin(-\pi u_1+\pi u_2+\pi {R_Y\over 2}z)}\right]\nonumber\\
&\times
\left[{\sin\left(-\pi u_2+\pi  \left( {R_Z\over2} -1\right)z\right)\over \sin(-\pi u_2+\pi {R_Z\over 2}z)}\right]\,
du_1\wedge du_2\wedge du_3.
\end{align}
It has poles at the hyperplanes
\begin{align}
\begin{split}
&H_X:~ u_1+{R_X\over 2}z=0,\\
&H_Y:~ -u_1 + u_2 +{R_Y\over 2}z=0,\\
&H_Z:~ - u_2 +{R_Z\over 2}z=0.
\end{split}
\end{align}
The corresponding charge covectors $Q_X,\,Q_Y,\, Q_Z$ on $\frak{h}^*\cong \mathbb{R}^2$ are shown in Figure \ref{triangle}. Since the theory is abelian, $\frak{g}_{u(1)}^* = \frak{h}^*$ and we can take $\zeta$ to be at any point on the plane $\frak{h}^*$. 

There are three chambers on the FI parameter space $\zeta$ in Figure \ref{triangle}. For a given $\zeta$, the index receives contributions from one of the three intersections $H_X\cap H_Y$, $H_Y\cap H_Z$, and $H_X\cap H_Z$, depending on which chamber $\zeta$ is in. A direct computation shows that all three chambers give the same answer. 

For example, if $\zeta \in \text{Cone}(Q_X,Q_Y)$, the Jeffrey-Kirwan residue operation is nonzero at $H_X\cap H_Y=\{u_1= - {R_X\over2}z,~u_2= -{R_X+R_Y\over 2}z\}$:
\begin{align}
\begin{split}
&\underset{u=H_X\cap H_Y}{\text{JK-Res}}\left( \{Q_X,Q_Y\}, \zeta \right)
{du_1\wedge du_2\over( u_1+{R_X\over2}z)(-u_1+u_2+{R_Y\over 2}z )}
\\
=& \left({1\over2\pi i}\right)^2 \oint_{u_1= -{R_X\over 2}z}  \oint_{u_2=u_1- {R_Y\over 2}z} \,
{du_1du_2\over  ( u_1+{R_X\over2}z)(-u_1+u_2+{R_Y\over 2}z ) }
\end{split}
\end{align}
The index can then be computed as
\begin{align}
\Omega(y,\zeta)&=\left[ - {\pi \over \sin(\pi z)}\right]^2
\oint_{u_1= -{R_X\over 2}z} {du_1\over 2\pi i} \oint_{u_2=u_1- {R_Y\over 2}z} {du_2\over2\pi i}\,
\left[   {\sin\left(\pi u_1+\pi  \left( {R_X\over2} -1\right)z\right)\over \sin(\pi u_1+ \pi {R_X\over 2}z)}  \right]\nonumber\\
&\times
\left[ {\sin\left(-\pi u_1+\pi u_2+\pi  \left( {R_Y\over2} -1\right)z\right)\over \sin(-\pi u_1+\pi u_2+\pi {R_Y\over 2}z)}\right]
\left[  {\sin\left(-\pi u_2+\pi  \left( {R_Z\over2} -1\right)z\right)\over \sin(-\pi u_2+\pi {R_Z\over 2}z)}  \right]\nonumber\\
&= { y^{ 1- {1\over k} }   -y^{-1 +{1\over k}}  \over y^{-{1\over k}} - y^{{1\over k}} }.
\end{align}

Note that the answer only depends on the sum of the $R$-charges $2/k$, but not on the individual assignments $R_X,\, R_Y,\, R_Z$. For $k=1$, $\Omega(y,\zeta)$ vanishes as expected. For $k>1$, however, the answer produced by the residue formula does not match that obtained by direct analysis. As explained above, this is no contradiction since in this case the superpotential does not satisfy our hypotheses.

\subsubsection{Generalized $XYZ$ Model}

\begin{figure}[here!]
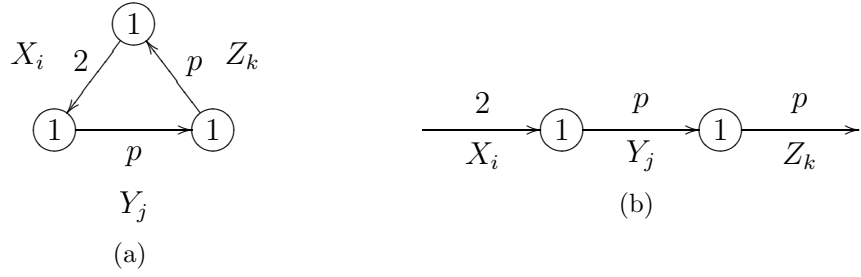

  \centering
\subfloat[]{\label{fig:2ppa}
\xy  0;<1pt,0pt>:<0pt,-1pt>::
(-300,0) *+{1}*\cir<8pt>{} ="1",
(-240,0) *+{1}*\cir<8pt>{} ="2",
(-270,-40) *+{1}*\cir<8pt>{} ="3",
(-270, 10) *+{p} ="b",
(-270, 29) *+{Y_j} ="B",
(-290, -26) *+{2} ="a",
(-310, -28) *+{X_i} ="A",
(-247, -24) *+{p} ="c",
(-229, -28) *+{Z_k} ="C",
\ar @{->} "1"; "2"
\ar @{->} "3"; "1"
\ar @{->} "2"; "3"
\endxy}
\hspace{.5in}
\subfloat[]{\label{fig:2ppb}
\xy  0;<1pt,0pt>:<0pt,-1pt>::
(-270, 10) *+{X_i} ="X",
(-270, -10) *+{2} ="x",
(-210, -10) *+{p} ="y",
(-210, 10) *+{Y_j} ="Y",
(-150, -10) *+{p} ="z",
(-150, 10) *+{Z_k} ="Z",
(-300,0) *+{\phantom{A}}="1",
(-240,0) *+{1}*\cir<8pt>{} ="2",
(-180,0) *+{1}*\cir<8pt>{} ="3",
(-120,0) *+{\phantom{B}}="4",
\ar @{->} "1"; "2",
\ar @{->} "2"; "3",
\ar @{->} "3"; "4",
\endxy}
  \caption{The generalized $XYZ$ model. The integers at the nodes denote the ranks of the associated gauge groups, while the integers at the arrows denote the number of bifundamental fields.  In (b), a choice of $U(1)$ decoupling.}
\end{figure}

We continue our study of models with superpotential but now with a more nontrivial index and wall-crossing phenomenon. 

The quiver diagram is shown in Figure \ref{fig:2ppa}. We have two $X_i$, $i=1,2$, and $p$ $Y_j$ and $Z_k$, $j,k=1,\cdots, p$, chiral multiplets. We will assume the $R$-charges for all the $X_i$ are the same and will be denoted by $R_X$. Similarly for $Y_j$ and $Z_k$. 

We will also assume $2=R_X+R_Y+R_Z$ so that the superpotential is cubic. The charge covectors are the same as the previous case shown in Figure \ref{triangle}. However, unlike the $XYZ$ model in the previous subsection, the index now does depend on the choice of the chamber.

In the $\zeta \in \text{Cone}(Q_X,Q_Y)$ chamber, the index is 
\begin{align}
\Omega(y,\zeta) &=\left[ - {\pi \over \sin(\pi z)}\right]^2
\oint_{u_1= -{R_X\over 2}z} {du_1\over 2\pi i} \oint_{u_2=u_1- {R_Y\over 2}z} {du_2\over2\pi i}\,
\left[{\sin\left(\pi u_1+\pi  \left( {R_X\over2} -1\right)z\right)\over \sin(\pi u_1+ \pi {R_X\over 2}z)}\right]^2\nonumber\\
&\times
\left[ {\sin\left(-\pi u_1+\pi u_2+\pi  \left( {R_Y\over2} -1\right)z\right)\over \sin(-\pi u_1+\pi u_2+\pi {R_Y\over 2}z)}\right]^p
\left[ {\sin\left(-\pi u_2+\pi  \left( {R_Z\over2} -1\right)z\right)\over \sin(-\pi u_2+\pi {R_Z\over 2}z)} \right]^p \nonumber\\
&=p.
\end{align}
Similarly the chamber $\zeta\in \text{Cone}(Q_X,Q_Z)$ gives the same answer as above.  

In the chamber $\zeta\in \text{Cone}(Q_Y, Q_Z)$, the index is
\begin{align}
\Omega(y,\zeta)&=
\left[ - {\pi \over \sin(\pi z)}\right]^2
\oint_{u_2= {R_Z\over 2}z} {du_2\over 2\pi i} \oint_{u_1=u_2+ {R_Y\over 2}z} {du_1\over2\pi i}\,
\left[ {\sin\left(\pi u_1+\pi  \left( {R_X\over2} -1\right)z\right)\over \sin(\pi u_1+ \pi {R_X\over 2}z)}\right]^2\nonumber\\
&\times
\left[ {\sin(-\pi u_1+\pi u_2+\pi  \left( {R_Y\over2} -1\right)z)\over \sin(-\pi u_1+\pi u_2+\pi {R_Y\over 2}z)}\right]^p
\left[{\sin(-\pi u_2+\pi  \left( {R_Z\over2} -1\right)z)\over \sin(-\pi u_2+\pi {R_Z\over 2}z)}\right]^p\nonumber\\
&=\begin{cases}
\sum_{ j=2}^p (j-1) \left(  y^{2(p-j)} + y^{-2(p-j)}\right) &\text{if}~p>1,\\
0&\text{if}~p=1. 
\end{cases}
\end{align}
Note that the index again does not depend on the individual $R$-charges $R_X,\, R_Y,\, R_Z$ but only on their sum. We assume $R_X+R_Y+R_Z=2$ to allow for the generic cubic superpotential.

These results match those obtained by directly quantizing the quiver moduli space in \cite{Cordova:2013bza}.

\subsubsection{4$d$ $\mathcal{N}=2$ $SU(3)$ Yang-Mills Theory}

\begin{figure}[here!]
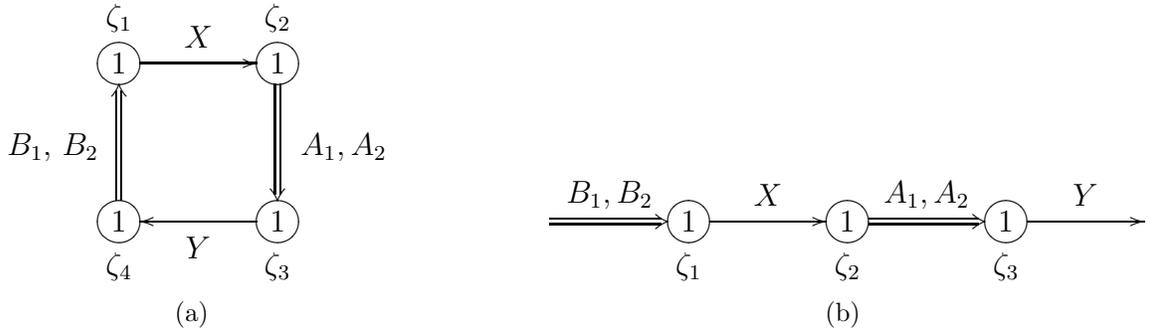

  \centering
\subfloat[]{\label{fig:SU(3)a}
\xy  0;<1pt,0pt>:<0pt,-1pt>::
(-300,17) *+{\zeta_4} ="q",
(-300,0) *+{1}*\cir<8pt>{} ="1",
(-240,0) *+{1}*\cir<8pt>{} ="2",
(-240,17) *+{\zeta_3} ="r",
(-300,-60) *+{1}*\cir<8pt>{} ="3",
(-300,-77) *+{\zeta_1} ="t",
(-240,-60) *+{1}*\cir<8pt>{} ="4",
(-240,-77) *+{\zeta_2} ="y",
(-270, 10) *+{Y} ="b",
(-270, -70) *+{X} ="d",
(-325, -28) *+{B_1,\, B_2} ="A",
(-215, -28) *+{A_1,A_2} ="C",
\ar @{<-} "1"; "2"
\ar @{<=} "3"; "1"
\ar @{->} "3";"4"
\ar @{=>} "4";"2"
\endxy}
\hspace{.5in}
\subfloat[]{\label{fig:SU(3)b}
\xy  0;<1pt,0pt>:<0pt,-1pt>::
(-270, -10) *+{B_1,B_2} ="x",
(-210, -10) *+{X} ="y",
(-150, -10) *+{A_1,A_2} ="z",
(-90, -10) *+{Y} ="z",
(-300,0) *+{\phantom{A}}="1",
(-240,17) *+{\zeta_1} ="a",
(-240,0) *+{1}*\cir<8pt>{} ="2",
(-180,0) *+{1}*\cir<8pt>{} ="3",
(-180,17) *+{\zeta_2} ="b",
(-120,0) *+{1}*\cir<8pt>{}="4",
(-120,17) *+{\zeta_3} ="a",
(-60,0) *+{\phantom{B}}="5",
\ar @{=>} "1"; "2",
\ar @{->} "2"; "3",
\ar @{=>} "3"; "4",
\ar @{->} "4";"5"
\endxy}
  \caption{The BPS quiver for 4$d$ $\mathcal{N}=2$ $SU(3)$ Yang-Mills theory.  The integers at the nodes denote the ranks of the associated gauge groups in the quantum mechanics, while the integers at the arrows denote the number of bifundamental fields. The $\zeta_{i}$ indicate our convention for FI parameters.  The corresponding BPS particle is a $W$-boson in the 4$d$ theory. In (b), a choice of $U(1)$ decoupling.}
\end{figure}

As a final example we consider the quiver quantum mechanics which governs the BPS states of four-dimensional $\mathcal{N}=2$  $SU(3)$ Yang-Mills theory \cite{Fiol:2000pd, Alim:2011kw} shown in Figure \ref{fig:SU(3)a}.  We study an example where the ranks of the quiver gauge groups are all one.  The corresponding BPS particle is a $W$-boson.  We expect that this particle is stable, and hence ground states of the quantum-mechanics exist, in the weak coupling region of the four-dimensional moduli space.  This is the region in $\zeta$-space where
\begin{equation}
 \zeta_{2}<\zeta_{3}, \hspace{.5in} \zeta_{4}<\zeta_{1}.\label{weakdef}
\end{equation}

The superpotential is
\begin{align}
\mathcal{W}= B_1  X A_1 Y - B_2 X A_2 Y.
\end{align}
From the symmetry of the quiver, we assume the $R$-charges for $X$ and $Y$ are the same and denote them by $R$. From the superpotential and the symmetry we deduce that the $R$-charges for $A_1,A_2,B_1,B_2$ are all $1-R$.

We decouple the $U(1)$ as in Figure \ref{fig:SU(3)b}. The meromorphic top form is
\begin{align}
&  Z_{1-\text{loop}} = \left[ - {\pi \over \sin (\pi z)  }\right]^3
\left[  {\sin\left( \pi u_1 +\pi \left( {1-R\over2} -1\right)z   \right) \over \sin\left( \pi u_1 + \pi {1-R\over 2} z\right)}  \right]^2
\left[  {\sin\left(- \pi u_1+\pi u_2 +\pi \left( {R\over2} -1\right)z   \right) \over \sin\left( - \pi u_1 +\pi u_2+ \pi {R\over 2} z\right)}  \right]\nonumber\\
&~~~~\times
\left[  {\sin\left(- \pi u_2+\pi u_3 +\pi\left( {1-R\over2}-1\right) z   \right) \over \sin\left(- \pi u_2+\pi u_3 + \pi {1-R\over 2} z\right)}  \right]^2
\left[  {\sin\left(- \pi u_3+\pi\left( {R\over2} -1\right)z  \right) \over \sin\left( - \pi u_3 + \pi {R\over 2} z\right)}  \right] 
du_1\wedge du_2\wedge du_3
\end{align}

There are four hyperplanes in $\frak{M}$ where $Z_{1-\text{loop}}$ has poles:
\begin{align}
\begin{split}
&H_1 :~ u_1 + {1-R\over 2}z=0,\\
&H_2 :~ -u_1 + u_2 + {R\over 2}z=0,\\
&H_3:~ -u_2+u_3 + {1-R\over2} z=0,\\
&H_4:~ -u_3 +{R\over 2}z=0.
\end{split}
\end{align}
The charge covectors in $\frak{h}^*$ that define these hyperplanes in $\frak{M}$ are
\begin{align}
\begin{split}
&Q_1 = (1,0,0),\\
&Q_2 = (-1,1,0),\\
&Q_3= (0,-1,1),\\
&Q_4= (0,0,-1).
\end{split}
\end{align}

$\frak{M}_{\text{sing}}^* $ contains the following four points from the intersections of any three of the four hyperplanes $H_i$:
\begin{align}
\begin{split}
&u^{(1)}_* = H_2\cap H_3\cap H_4= \left(  {1+R\over 2}z  ,\, {1\over 2}z  , \,  {R\over 2} z  \right), \\
&u^{(2)}_* = H_1\cap H_3\cap H_4= \left(  - {1-R\over 2}z ,\, {1\over 2}z,  \,{R\over 2} z   \right), \\
&u^{(3)}_* = H_1\cap H_2\cap H_4 = \left( -{1-R\over2} z, \, -{1\over 2}z, \,{R\over 2}z   \right),\\
&u^{(4)}_* = H_1\cap H_2\cap H_3 = \left(-{1-R\over 2}z , \,- {1\over 2}z ,\,  -{2-R\over 2}z  \right).
\end{split}
\end{align}
Given the FI parameter $\zeta\in \frak{h}^*$, it belongs to a cone generated by three of the four $Q_i$'s.  The Jeffrey-Kirwan residue only receives contribution from the intersection of the three corresponding hyperplanes $H_i$. For example, we can write the Jeffrey-Kirwan residue operation in the Cone$(Q_2,Q_3,Q_4)$ chamber as
\begin{align}
\begin{split}
&\underset{u=u_*^{(1)} }{\text{JK-Res} }\left(
\{Q_2,Q_3,Q_4\} , \zeta   \right)
{du_1\wedge du_2\wedge du_3 \over
\left( - u_1 +u_2 +{R\over2}z \right)\left(-u_2 +u_3 +{1-R\over 2}z \right) \left( -u_3 +{R\over 2}z  \right)
}  \\
=& 
(-1)^3\left({1\over 2\pi i}\right)^3 
\oint_{u_3= {R\over 2} z}du_3
\oint_{u_2= u_3+ {1-R\over 2}z } du_2
  \oint_{u_1=u_2+{R\over 2}z} {du_1}\\
  \times&
  {1\over \left( - u_1 +u_2 +{R\over2}z \right)\left(-u_2 +u_3 +{1-R\over 2}z \right) \left( -u_3 +{R\over 2}z \right)
}.
\end{split}
\end{align}

The index in the four chambers are 
\begin{align}
\Omega(y,\zeta) = \begin{cases}
0 &\text{if}~~  \zeta \in \text{Cone}(Q_2,Q_3,Q_4),\\
y+y^{-1} &\text{if}~ ~ \zeta \in \text{Cone}(Q_1,Q_3,Q_4),\\
0 &\text{if}~~   \zeta \in \text{Cone}(Q_1,Q_2,Q_4),\\
y+y^{-1} &\text{if}~~   \zeta \in \text{Cone}(Q_1,Q_2,Q_3).
\end{cases}
\end{align}
In terms of the components of the FI parameters, the chambers can be described as
\begin{align}
\Omega(y,\zeta) = \begin{cases}
0 &\text{if}~~  \zeta_4>0,~\zeta_1<0,~\zeta_1+\zeta_2<0,\\
y+y^{-1} &\text{if}~~ \zeta_1>0,~\zeta_2<0,~\zeta_2+\zeta_3<0,\\
0 &\text{if}~ ~ \zeta_2>0,~\zeta_3<0,~\zeta_1+\zeta_2>0,\\
y+y^{-1} &\text{if}~~   \zeta_3>0,~\zeta_4<0,~\zeta_2+\zeta_3>0.
\end{cases}
\end{align}

This agrees with the expectation \eqref{weakdef} and provides a complete picture of the walls of marginal stability where the $W$-boson decays.

\section*{Acknowledgements} 
The work of CC is support by a Junior Fellowship at the Harvard Society of Fellows.  The work of SHS is supported by the Kao Fellowship and the An Wang Fellowship at Harvard University. 
\newpage

\bibliography{SQMIndexv2}
\bibliographystyle{utphys}
\end{document}